\documentclass[10pt]{article}

\usepackage[superscript,sort]{cite}

\usepackage{ifthen,ifpdf}
\ifpdf
	\usepackage[pdftex]{hyperref}\hypersetup{colorlinks=true,linkcolor=black,citecolor=black,urlcolor=blue,backref=page,bookmarks=true,breaklinks=true,plainpages=false }
	\usepackage[pdftex]{graphicx}
	\usepackage[usenames,pdftex]{color}
\else
	\usepackage[colorlinks=true,breaklinks=true,plainpages=false]{hyperref}
	\usepackage{graphicx}
	\usepackage[usenames]{color}
\fi

\usepackage{amsthm,amscd,amsxtra,amsfonts,amsmath,amssymb,multirow,dsfont}
\usepackage{wrapfig}
\usepackage[footnotesize]{caption}
\usepackage[tiny,compact]{titlesec}
\usepackage[textwidth=0.8in,textsize=footnotesize]{todonotes}
\usepackage{algorithm,algorithmic,extarrows}

\begin{document}

\title{Persistent spectral based machine learning (PerSpect ML) for drug design
}

\author{
Zhenyu Meng$^1$ and
Kelin Xia$^{1,2}$ \footnote{ Address correspondences  to Kelin Xia. E-mail:xiakelin@ntu.edu.sg}\\
$^1$Division of Mathematical Sciences, School of Physical and Mathematical Sciences, \\
Nanyang Technological University, Singapore 637371\\
$^2$  School of Biological Sciences, \\
Nanyang Technological University, Singapore 637371\\
}

\date{\today}
\maketitle

\begin{abstract}
In this paper, we propose persistent spectral based machine learning (PerSpect ML) models for drug design. Persistent spectral models, including persistent spectral graph, persistent spectral simplicial complex and persistent spectral hypergraph, are proposed based on spectral graph theory, spectral simplicial complex theory and spectral hypergraph theory, respectively. Different from all previous spectral models, a filtration process, as proposed in persistent homology, is introduced to generate multiscale spectral models. More specifically, from the filtration process, a series of nested topological representations, i,e., graphs, simplicial complexes, and hypergraphs, can be systematically generated and their spectral information can be obtained. Persistent spectral variables are defined as the function of spectral variables over the filtration value. Mathematically, persistent multiplicity (of zero eigenvalues) is exactly the persistent Betti number (or Betti curve). We consider 11 persistent spectral variables and use them as the feature for machine learning models in protein-ligand binding affinity prediction. We systematically test our models on three most commonly-used databases, including PDBbind-2007, PDBbind-2013 and PDBbind-2016. Our results, for all these databases, are better than all existing models, as far as we know. This demonstrates the great power of our PerSpect ML in molecular data analysis and drug design.
\end{abstract}


\newpage

{\setcounter{tocdepth}{5} \tableofcontents}

\newpage

\section{Introduction}

Data-driven learning models are among the most important and rapidly evolving areas in chemoinformatics and bioinformatics \cite{lo2018machine,puzyn2010recent}. Greatly benefit from the accumulation of experimental data, machine learning and deep learning models have contributed significantly to various aspects of virtual screening in drug design. In particular, machine-learning-based scoring functions have dramatically increased the accuracy of binding affinity prediction and delivered better results than traditional physics-based, knowledge-based and empirical-based models \cite{li2015improving}. Featurization or feature engineering is key to the performance of machine learning and deep learning models in biomolecular systems. To characterize the structural, physical, chemical, and biological properties, more than 5000 molecular descriptors and chemical descriptors are proposed \cite{puzyn2010recent,lo2018machine}. These descriptors cover information from molecular formula, fragments, motifs, topological features, geometric features, conformation properties, hydrophobicity, electronic properties, steric properties, etc. They are widely-used in quantitative structure-activity relationship (QSAR) and quantitative structure-property relationship (QSPR) models. More importantly, these descriptors can be combined to form a fixed-length vector, known as molecular fingerprint. These equal-sized molecular fingerprints are representations of molecules and can be used as input features for machine learning models. Various softwares, such as RDkit \cite{landrum2006rdkit}, Open babel \cite{oboyle2011open}, ChemoPy \cite{cao2013chemopy}, etc, are built for the automatical generation of these molecular descriptors.

Recently, advanced mathematics models, including algebraic topology, differential geometry and algebraic graph theory, are proposed for the representation and featurization of biomolecular systems and can significantly enhance the performance of learning models in drug design \cite{wei2017persistent,wei2017mathematics,nguyen2020review,cang:2018representability}. Different from other molecular descriptors, three unique kinds of invariants, i.e., topological invariant (Betti numbers), geometric invariant (curvatures) and algebraic graph invariant (eigenvalues), are considered. The combination of these invariants with learning models has achieved great successes in various aspects of drug design, including protein-ligand binding affinity prediction \cite{cang:2017topologynet,cang:2017integration,nguyen:2017rigidity,cang2018integration,nguyen2019agl}, protein stability change upon mutation prediction \cite{cang:2017analysis,cang:2018representability}, toxicity prediction \cite{wu:2018quantitative}, solvation free energy prediction \cite{wang2016automatic,wang2018breaking}, partition coefficient and aqueous solubility \cite{wu2018topp}, binding pocket detection \cite{zhao2018protein}, and drug discovery \cite{grow2019generative}. More interestingly, these advanced-mathematics-based machine learning models have constantly achieved some of the best results in D3R Grand challenge \cite{nguyen2018mathematical,nguyen2019mathdl,nguyen2019mathematical}.

Motivated by the great success of these advance mathematics models in drug design, we propose persistent spectral (PerSpect) theory and persistent spectral based machine learning (PerSpect ML). Our persistent spectral theory cover three basic models, including PerSpect graph \cite{wang2019persistent}, PerSpect simplicial complex and PerSpect hypergraph. Mathematically, graph, simplicial complex and hypergraph and three topological models for structure characterization. Based on them, spectral graph theory \cite{ChungOverview,spielman2007spectral}, spectral simplicial complex \cite{eckmann1944harmonische,muhammad2006control,horak2013spectra,barbarossa2019topological} and spectral hypergraph \cite{feng1996spectra,sun2008hypergraph,cooper2012spectra,lu2011high,barbarossa2016introduction} are proposed. In spectral graph models, Laplacian matrixes are proposed as the algebraic description of graphs. The eigen spectral information of the Laplacian matrix, including Fiedler value, Cheeger constant, vertex and edge expansion, graph heat kernel and flow, etc, can then be used in characterization of graph properties \cite{ChungOverview,spielman2007spectral}. In spectral simplicial complex, combinatorial Laplacians or Hodge Laplacians can be defined from boundary matrixes, which characterize the topological connection between low-dimensional simplexes and high-dimensional simplexes. Essentially, graph Laplacian describes the relation between 1-simplexes (edges) and 0-simplexes (vertices), while combinatorial Laplacians are the generalization of the relation to higher-dimensional simplexes, such as, 2-simplexes (triangles), 3-simplexes (tetrahedrons), etc.  
In spectral hypergraph, boundary matrix or incident matrix can be defined between hyperedges and vertices. Hypergraph Laplacian matrix can then be constructed from the incident matrix. Hypergraph Laplacians can also be defined as combinatorial Laplacians of the Clique complex, which is induced from the hypergraph.

Different from all previous spectral models, persistent spectral theory describe structures from not one but multiple different scales.  This multiscale representation is achieved through a filtration process, which is the key component of persistent homology \cite{Edelsbrunner:2002,Zomorodian:2005}. During the filtration process, a nested sequence of topological structures, which can be graphs, simplicial complexes, or hypergraphs, can be systematically generated. Their spectral properties, which are changed with filtration values, are defined as PerSpect variables. These PerSpect variables characterize not only the global topological information, but also the geometric information that directly related to topological variations, in a way similar as persistent homology. Therefore, PerSpect variables can be used in both qualitative description and quantitative characterization. In particular, persistent multiplicity (of zero eigenvalues) is exactly persistent Betti number. This means that PerSpect theory incorporates persistent homology information. Moreover, PerSpect variables can be discretized into feature vectors for various learning models, such as support vector machine, random forest, gradient boost tree, neural network, convolution neural network, etc. The unique multiscale properties of these features, which balance structure complexity and data simplification, enables a better structure representation and can boost the performance of learning models.

In this paper, we consider PerSpect simplicial complex based ML model for protein-ligand binding affinity prediction in drug design. We train our model on three widely-used protein-ligand databases \cite{PDBBind:2015}, including PDBbind-2007, PDBbind-2013 and PDBbind-2016. Similar to existing models, Pearson correlation coefficient (PCC) and root-mean square error (RMSE) are used as measurements for the performance of the model. We systematically compare our prediction results with all the state-of-art results in protein-ligand binding affinity prediction, as far as we known.  It has been found that our PerSpect ML model has delivered the best results in terms of both PCC and RMSE in all the three test cases \cite{liu2015classification,li2015improving,feinberg2018potentialnet,wojcikowski2019development,jimenez2018k,stepniewska2018development,su2018comparative,zheng2019onionnet,afifi2018improving,feinberg2018potentialnet,boyles2019learning}.

\section{Theory and methods}

\subsection{Topological representations}\label{sec:Flexibility}

\begin{figure}
	\begin{center}
		\begin{tabular}{c}
			\includegraphics[width=0.7\textwidth]{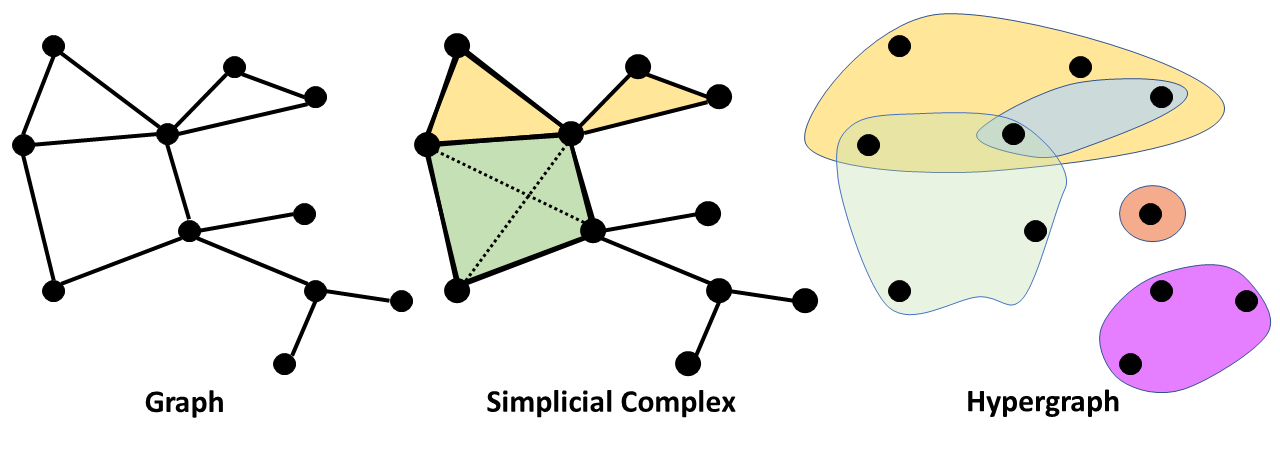}
		\end{tabular}
	\end{center}
	\caption{The illustration of the three topological representations, i.e., graph, simplicial complex and hypergraph. Mathematically, a graph is a simplicial complex with only vertices (0-simplexes) and edges (1-simplexes). The simplicial complex is a generalization of graphs into their higher-dimensional counterparts. Hypergraph is a further generalization of simplicial complex by the replacement of simplexes with hyperedges.
	}
	\label{fig:Top_Representation}
\end{figure}

\paragraph{Graph}

Graph or network models have been applied to various material, chemical and biological systems. In these models, atoms and bonds are usually simplified as vertices and edges. Mathematically, a graph representation can be denoted as $G(V,E)$, where $V= \{v_i;i=1,2,...,N \}$ are vertex set with $N=|V|$ the total number. Here $E=\{e_i=(v_{i_1},v_{i_2});1\leq i_1 \leq N, 1\leq i_2 \leq N \}$ denotes the edge set.

\paragraph{Simplical complex}
A simplicial complex is the generalization of a graph into its higher-dimensional counterpart.  The simplicial complex is composed of simplexes. Each simplex is a finite set of vertices, and can be viewed geometrically as, a point (0-simplex), an edge (1-simplex), a triangle (2-simplex), a tetrahedron (3-simplex), and their k-dimensional counterpart (k-simplex). More specifically, a $k$-simplex $\sigma^k=\{v_0,v_1,v_2,\cdots,v_k\}$ is the convex hull formed by $k+1$ affinely independent points $v_0,v_1,v_2,\cdots,v_k$ as follows,
\begin{eqnarray}\label{eq:couple_matrix1}\nonumber
\sigma^k=\left\{\lambda_0 v_0+\lambda_1 v_1+ \cdots +\lambda_k v_k \mid \sum^{k}_{i=0}\lambda_i=1;\forall i, 0\leq \lambda_i \leq 1 \right\}.
\end{eqnarray}
The $i$-th dimensional face of $\sigma^k$ ($i<k$) is the convex hull formed by $i+1$ vertices from the set of $k+1$ points $v_0, v_1, v_2, \cdots, v_k$. The simplexes are the basic components for a simplicial complex.

A simplicial complex $K$ is a finite set of simplexes that satisfy two conditions. Firstly, any face of a simplex from  $K$  is also in  $K$. Secondly, the intersection of any two simplexes in $K$ is either empty or a shared face. A $k$-th chain group $C_k$ is an Abelian group generated by oriented $k$-simplexes $\sigma^k$, which are simplexes together with an orientation, i.e., ordering of their vertex set. The boundary operator $\partial_k$ ($C_k \rightarrow C_{k-1}$) for an oriented $k$-simplex $\sigma^k$ can be denoted as,
\begin{eqnarray}\nonumber
\partial_k \sigma^k = \sum^{k}_{i=0} (-1)^i[ v_0, v_1, v_2, \cdots, \hat{v_i}, \cdots, v_k ].
\end{eqnarray}
Here $[v_0, v_1, v_2, \cdots ,\hat{v_i}, \cdots, v_k ]$ is a oriented $(k-1)$-simplex, which is generated by the original set of vertices except $v_i$. The boundary operator maps a simplex to its faces and it guarantees that  $\partial_{k-1}\partial_k=0$.

To facilitate a better description, we use notation $\sigma_j^{k-1} \subset \sigma_i^{k}$ to indicate that   $\sigma_j^{k-1}$ is a face of $\sigma_i^{k}$, and notation $\sigma_j^{k-1} \sim \sigma_i^{k}$ if they have the same orientation, i.e., oriented similarly. For two oriented $k$-simplexes, $\sigma_i^k$ and $\sigma_j^k$, of a simplicial complex $K$, they are upper adjacent, denoted as  $\sigma_i^k \frown \sigma_j^k$, if they are faces of a common $(k+1)$-simplex; they are lower adjacent, denoted as $\sigma_i^k \smile \sigma_j^k$, if they share a common $(k-1)$-simplex as they face. Moreover, if the orientations (or signs) of their common lower simplex are the same, it is called similar common lower simplex ($\sigma_i^k \smile \sigma_j^k$ and $\sigma_i^k \sim  \sigma_j^k$); if their orientations are differently, it is called dissimilar common lower simplex ($\sigma_i^k \smile \sigma_j^k$ and $\sigma_i^k \not \sim  \sigma_j^k$). The (upper) degree of a $k$-simplex $\sigma_i^k$, denoted as $d(\sigma^k)$, is the number of $(k+1)$-simplexes, of which $\sigma_i^k$ is a face.

\paragraph{Hypergraph}
A hypergraph is a generalization of graph in which an edge is made of a set of vertices. Mathematically, a hypergraph $H(V,E^h)$ consists of a set of vertices (denoted as $V$), and a set of hyperedges (denoted as $E^h$). Each hyperedge contains an arbitrarily number of vertices, and can be regarded as a subset of $V$. A hyperedge $e_i^h$ is said to be incident with a vertex $v_j$, when the vertice is in the hyperedge, i.e., $v_j \in e_i^h$. Note that a hypergraph can also be viewed as a generalization of simplicial complex.

An illustration of graph, simplicial complex and hypergraph can be found in Figure \ref{fig:Top_Representation}. These topological representations are made from the same set of vertices, but they characterize different topological connections.

\subsection{Spectral theories}
A systematic characterization, identification, comparison, and analysis of structure data, from material, chemical and biological systems, are usually complicated due to their high dimensionality and complexity. Spectral graph theory is proposed to reduce the data dimensionality and complexity by studying the spectral information of connectivity matrixes, constructed from the structure data. These connectivity matrixes include incidence matrix, adjacency matrix, (normalized) Laplacian matrix, Hessian matrix, etc. Spectral information includes eigenvalues, eigenvectors, eigenfunctions, and other related properties, such as, Cheeger constant, edge expansion, vertex expansion, graph flow, graph random walk, heat kernel of graph, etc. Mathematically, spectral graph theory can be generalized into spectral simplicial complex and spectral hypergraph.

\paragraph{Spectral graph}
In spectral graph theory, a graph $G(V,E)$ is represented by its adjacency matrix and Laplacian matrix \cite{ChungOverview, spielman2007spectral, Mohar:1991laplacian, von:2007tutorial}. The adjacency matrix ${\bf A}$ describes the connectivity information and can be expressed as,
\begin{eqnarray}\label{eq:graph_adjacency}\nonumber
A(i,j)=\begin{cases} \begin{array}{ll}
	        1, & (v_i,v_j) \in E\\
            0, & (v_i,v_j) \not \in E.\\
	      \end{array}
\end{cases}
\end{eqnarray}
The degree of a vertex $v_i$ is the total number of edges that are connected to vertex $v_i$, i.e., $d(v_i)=\sum_{i \neq j}^N A(i,j)$. The vertex diagonal matrix ${\bf D}$ can be defined as,
\begin{eqnarray}\label{eq:graph_diagonal}
D(i,j)=\begin{cases} \begin{array}{ll} \nonumber
	        \sum_{i \neq j}^N A(i,j), & i=j\\
            0, & i \neq j.
	      \end{array}
\end{cases}
\end{eqnarray}
Laplacian matrix, also known as admittance matrix and Kirchhoff matrix, is defined as ${\bf L}={\bf D}-{\bf A}$. More specifically, it can be expressed as,
\begin{eqnarray}\label{eq:graph_Laplacian}
L(i,j)=\begin{cases} \begin{array}{ll}
			d(v_i), &i=j\\
            -1, & i \neq j~{\rm and} ~ (v_i,v_j) \in E\\
            0,  &i \neq j~{\rm and} ~ (v_i,v_j) \not \in E.
	      \end{array}
\end{cases}
\end{eqnarray}

The Laplacian matrix has many important properties. It is always positive-semidefinite, thus all its eigenvalues are non-negative. In particular, the number (multiplicity) of zero eigenvalues equals to an topological invariant, known as $\beta_0$, which counts the number of connected components in the graph. The second smallest eigenvalue, i.e., the first non-zero eigenvalue, is called Fiedler value or algebraic connectivity, which describes the general connectivity of the graph. The corresponding eigenvector can be used to subdivide the graph into two well-connected subgraphs. All eigenvalues and eigenvectors form an eigen spectrum and spectral graph theory studies the properties of the graph eigen spectrum.

There are two types of normalized Laplacian matrixes, including the symmetric normalized Laplacian matrix, which is defined as ${\bf L}_{\text{sym}}={\bf D}^{-1/2}{\bf L}{\bf D}^{-1/2}$, and random walk normalized Laplacian, which is defined as ${\bf L}_{\text{rw}}={\bf D}^{-}{\bf L}$.

\paragraph{Spectral simplicial complex}

The spectral simplicial complex theory studies the spectral properties of combinatorial Laplacian (or Hodge Laplacian) matrixes, that are constructed based on a simplicial complex \cite{eckmann1944harmonische,muhammad2006control,horak2013spectra,barbarossa2019topological,mukherjee2016random,parzanchevski2017simplicial,shukla2020spectral,torres2020simplicial}.

For an oriented simplicial complex, its $k$-th boundary (or incidence) matrix ${\bf B}_k$ can be defined as follows,
\begin{eqnarray}\label{eq:Boundary} \nonumber
B_k(i,j)=\left\{\begin{array}{ll}
1, &\text{if } \sigma_i^{k-1} \subset \sigma_j^{k} ~\text{and }~ \sigma_i^{k-1} \sim \sigma_j^{k}  \\
-1, &\text{if } \sigma_i^{k-1} \subset \sigma_j^{k} ~\text{and }~ \sigma_i^{k-1} \not\sim \sigma_j^{k}  \\
0, &\text{if } \sigma_i^{k-1} \not\subset \sigma_j^{k}.
\end{array}
\right.
\end{eqnarray}

These boundary matrixes satisfy the condition that ${\bf B}_k{\bf B}_{k+1}={\bf 0}$. The $k$-th combinatorial Laplacian matrix can be expressed as follows,
\begin{eqnarray}\label{eq:Simplex_Laplacian} \nonumber
{\bf L}_k={\bf B}^T_k{\bf B}_{k}+{\bf B}_{k+1}{\bf B}^T_{k+1}.
\end{eqnarray}
Note that $0$-th combinatorial Laplacian is,
 $${\bf L}_0={\bf B}_1{\bf B}^T_{1}.$$
Further, if the highest order of the simplicial complex $K$ is $n$, then the $n$-th combinatorial Laplacian matrix is ${\bf L}_n={\bf B}^T_n{\bf B}_{n}$.

The above combinatorial Laplacian matrixes can be explicitly described in terms of the simplex relations.  More specifically, ${\bf L}_0$, i.e., when $k=0$, can be expressed as,
\begin{eqnarray}\label{eq:Simplex_Laplacian_dim0}
L_0(i,j)=\left\{\begin{array}{ll}
d(\sigma_i^{0}), &\text{if } i=j \\
-1, & \text{if } i \neq j~\text{and } \sigma_i^{0} \frown \sigma_j^{0} \\
0, & \text{if } i \neq j~\text{and } \sigma_i^{0} \not \frown \sigma_j^{0}.
\end{array}
\right.
\end{eqnarray}
It can be seen that this expression is exactly the graph Laplacian as in Eq. (\ref{eq:graph_Laplacian}). Further, when $k>0$, ${\bf L}_k$ can be expressed as \cite{muhammad2006control},
\begin{eqnarray}\label{eq:Simplex_Laplacian_long}
L_k(i,j)=\left\{\begin{array}{l}
d(\sigma_i^{k})+ k+1, \quad \text{if } i=j \\
1, \qquad \text{if } i \neq j, \sigma_i^{k} \not \frown\sigma_j^{k}, \sigma_i^{k} \smile \sigma_j^{k}~\text{and } \sigma_i^k \sim  \sigma_j^k   \\
-1, \quad \text{if } i \neq j, \sigma_i^{k} \not \frown \sigma_j^{k}, \sigma_i^{k} \smile \sigma_j^{k}~\text{and } \sigma_i^k \not \sim  \sigma_j^k \\
0, \qquad \text{if } i \neq j, \sigma_i^{k} \frown \sigma_j^{k} ~\text{or }~ \sigma_i^{k} \not \smile \sigma_j^{k}.
\end{array}
\right.
\end{eqnarray}
The eigenvalues of combinatorial Laplacian matrixes are independent of the choice of the orientation \cite{horak2013spectra}. Further, the multiplicity of zero eigenvalues, i.e., the total number of zero eigenvalues, of ${\bf L}_k$ equals to the $k$-th Betti number $\beta_k$. Geometrically, $\beta_0$ is the number of connected components,  $\beta_1$ is the number of circles or loops, and $\beta_2$ is the number of voids or cavities.

We can define the $k$-th combinatorial down Laplacian matrix as ${\bf L}^\text{down}_k={\bf B}^T_k{\bf B}_{k}$ and combinatorial up Laplacian matrix as ${\bf L}^\text{up}_k={\bf B}_{k+1}{\bf B}^T_{k+1}$. These matrixes have very interesting spectral properties \cite{barbarossa2019topological}. Firstly, eigenvectors associated nonzero eigenvalues of ${\bf L}^\text{down}_k$ are orthogonal to eigenvectors from nonzero eigenvalues of ${\bf L}^\text{up}_k$; Secondly, nonzero eigenvalues of of ${\bf L}_k$ are either the eigenvalues of  ${\bf L}^\text{down}_k$ or those of ${\bf L}^\text{up}_k$; Thirdly, eigenvectors associated with nonzero eigenvalues of ${\bf L}_k$ are either eigenvectors of  ${\bf L}^\text{down}_k$ or those of ${\bf L}^\text{up}_k$.

\paragraph{Spectral hypergraph}
Laplacian matrixes can also be defined on hypergraph \cite{feng1996spectra,sun2008hypergraph,cooper2012spectra,lu2011high,barbarossa2016introduction}. One way to do that is to employ a clique expansion, in which a graph is construct from a hypergraph $H(V,E)$ by replacing each hyperedge with an edge for each pair of vertices in this hyperedge. A graph Laplacian matrix can then be defined on this hypergraph-induced graph. Note that the clique expansion also generate a clique complex, and combinatorial Laplacian matrixes can also be constructed based on it.

The other way is to directly use the incidence matrix. In a hypergraph, an incidence matrix $\bf H$ can be defined as follows,
\begin{eqnarray}\label{eq:Laplacian_element} \nonumber
H(i,j)=\left\{\begin{array}{ll}
1, &\text{if } v_i \in e^h_j\\
0, &\text{if } v_i \not \in e^h_j.
\end{array}
\right.
\end{eqnarray}
The vertex diagonal matrix ${\bf D}_v$ is,
\begin{eqnarray}\label{eq:couple_matrix27}\nonumber
D_v(i,j)=\begin{cases} \begin{array}{ll}
	        \sum_{j} H(i,j), & i=j\\
            0, & i \neq j.
	      \end{array}
\end{cases}
\end{eqnarray}
The hypergraph adjacent matrix is then defined as ${\bf A}={\bf H}{\bf H}^T-{\bf D}_v$. And the unnormalized hypergraph Laplacian matrix is defined as,
\begin{eqnarray}\label{eq:hypergraph_Laplacian} \nonumber
{\bf L}=2{\bf D}_v-{\bf H}{\bf H}^T.
\end{eqnarray}
Similar to the graph models, the symmetric normalized hypergraph Laplacian is defined as ${\bf L}_{\text{sym}}=2{\bf I}-{\bf D}^{-1/2}_v{\bf H}{\bf H}^T{\bf D}^{-1/2}_v$ with ${\bf I}$ the identity matrix.  The random walk hypergraph Laplacian is defined as ${\bf L}_{\text{rw}}=2{\bf I}-{\bf D}^{-1}_v{\bf H}{\bf H}^T$.

\subsection{Persistent spectral theory}

\begin{figure}
	\begin{center}
		\begin{tabular}{c}
			\includegraphics[width=0.9\textwidth]{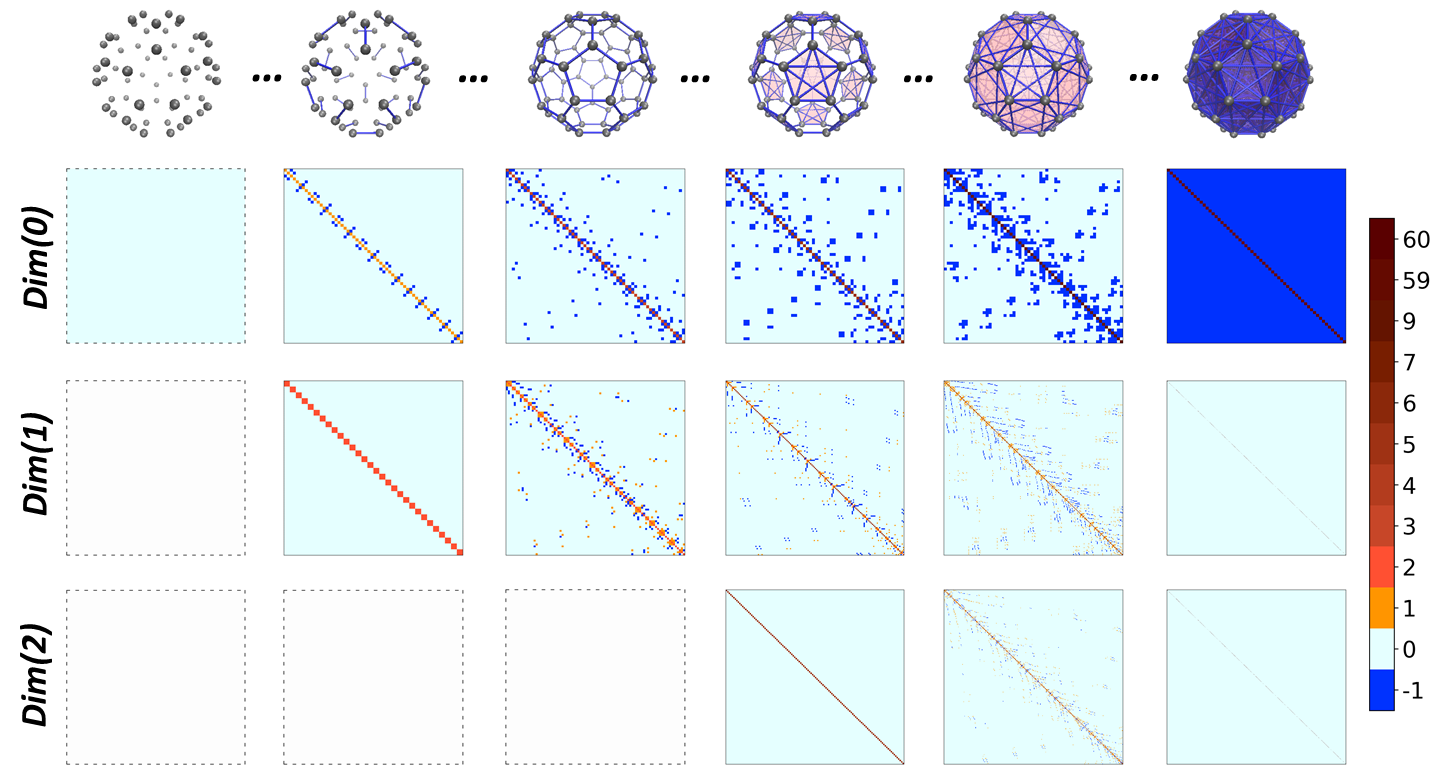}
		\end{tabular}
	\end{center}
	\caption{Persistent combinatorial Laplacian matrixes for simplicial complexes from a filtration process of fullerene C$_{60}$. Only combinatorial Laplacian matrices at $\text{Dim}(0)$ to $\text{Dim}(2)$ are illustrated. During the filtration, $\text{Dim}(0)$ Laplacian matrix changes from all-zero-entry matrix, meaning no connection at all, to a matrix with all offdiagonal entries as -1, representing a complete graph. For $\text{Dim}(1)$ and $\text{Dim}(2)$ Laplacian matrices, the number of their offdiagonal non-zero entries increases at early state of filtration, then systematically decreases and goes to zero, resulting in two diagonal matrices.
	}
	\label{fig:c60_matrix}
\end{figure}

\paragraph{Filtration}
A filtration process naturally generates a mutliscale representation \cite{Edelsbrunner:2002}. Filtration parameter, denoted as $f$ and key to the filtration process, is usually chosen as sphere radius (or diameter) for point cloud data, edge weight for graphs, and isovalue (or level set value) for density data. A systematical increase (or decrease) of the value for the filtration parameter will induce a sequence of hierarchical topological representations, which can be not only simplicial complexes, but also graphs and hypergraphs. For instance, a filtration operation on a distance matrix, i.e., a matrix with distances between any two vertices as its entries, can be defined by using a cutoff value as the filtration parameter. More specifically, if the distance between two vertices is smaller than the cutoff value, an edge is formed between them. In this way, a systematical increase (or decrease) of the cutoff value will deliver a series of nested graphs, with the graph produced at a lower cutoff value as a part (or a subset) of the graph produced at a larger cutoff value. Similarly, nested simplicial complexes can be constructed by using various definitions of complexes, such as Vietoris-Rips complex, $\check{C}$ech complex, Alpha complex, cubical complex, Morse complex, etc. Nested hypergraphs can also be generated by using a suitable definition of hyperedge.

\paragraph{Persistent spectral theory}
The essential idea of our PerSpect theory is to provide a new mathematical representation that characterize the intrinsic topological and geometric information of the data. Different from all previous spectral models, our PerSpect theory considers not the eigen spectrum information of the graph, simplicial complex or hypergraph, constructed from a data at a particular scale, instead they focus on the variation of the eigen spectrum of these topological representations during a filtration process. Stated differently, our PerSpect theory studies the change of eigen spectrum in an ``evolution" process, during which the structure of graph, simplicial complex or hypergraph ``evolves" from a set of isolated vertices to a fully-connected topology, according to their inner structure connectivity and a predefined filtration parameter.

Mathematically, a filtration operation will deliver a nested sequence of graphs as follows,
\begin{eqnarray} \nonumber
 G^0 \subseteq G^1 \subseteq \cdots \subseteq G^m.
\end{eqnarray}
Here $i$-th graph $G^i$ is generated at a certain filtration value $f_i$. Computationally, we can equally divide the filtration region (of the filtration parameter) into $m$ intervals and consider topological representations at each interval. A series of Laplacian matrixes $\{{\bf L}^i|_{i=1,2,...,m} \}$ can be generated from these graphs. 
Further, a nested sequence of simplicial complexes can also be generated from a filtration process,
\begin{eqnarray} \nonumber
 K^0 \subseteq K^1 \subseteq \cdots \subseteq K^m.
\end{eqnarray}
Similarly,  the $i$-th simplicial complex $K^i$ is generated at filtration value $f_i$. Combinatorial Laplacian matrix series $\{{\bf L}^i_k|_{i=1,2,...,m; k=0,1,2,...} \}$ can be constructed from these simplicial complexes. Note that the size of these Laplacian matrixes may be different. Moreover, with a suitable filtration process, a nested sequence of hypergraph can be generated as follows,
\begin{eqnarray} \nonumber
 H^0 \subseteq H^1 \subseteq \cdots \subseteq H^m.
\end{eqnarray}
Hypergraph Laplacian matrix series $\{{\bf L}^i|_{i=1,2,...,m} \}$ can be constructed accordingly.

PerSpect theory studies the variation of the spectral information from the series of Laplacian matrixes. PerSpect variables and functions can be defined on the series of eigen spectrums. These PerSpect variables can incorporate both geometric and topological information in them. For instance, the multiplicity (or number) of $\text{Dim}(k)$ zero eigenvalues equals to Betti number $\beta_k$, thus persistent multiplicity, which is defined as the multiplicity of $\text{Dim}(k)$ zero eigenvalues over a filtration process, is exactly the Persistent Betti number or Betti curve. Further, we can consider the basic statistic properties, such as mean, standard deviation, maximum and minimum, of all non-zero eigenvalues, and define four other PerSpect variables, i.e., persistent mean, persistent standard deviation, persistent maximum and persistent minimum. Other spectral information, including algebraic connectivity, modularity, Cheeger constant, vertex/edge expansion, and other flow, random walk, and heat kernel related properties, can also be generalized into their corresponding PerSpect variables or functions.

\begin{figure}
	\begin{center}
		\begin{tabular}{c}
			\includegraphics[width=0.8\textwidth]{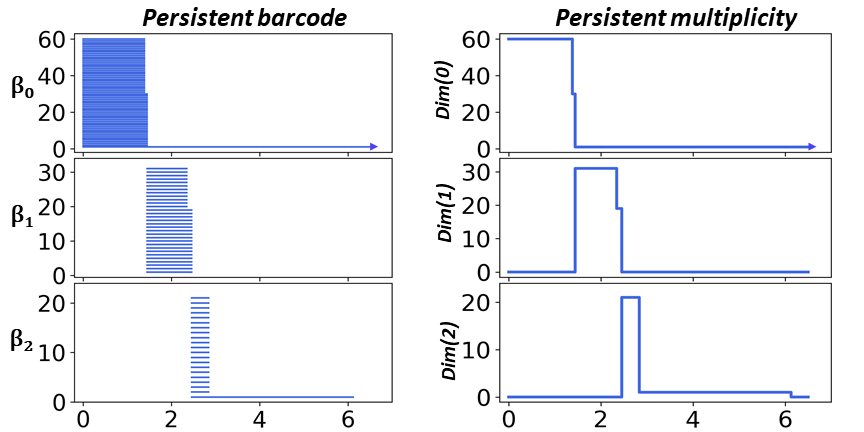}
		\end{tabular}
	\end{center}
	\caption{The comparison of persistent barcodes and persistent multiplicities of fullerene C$_{60}$. The $\text{Dim}(k)$ persistent multiplicity is multiplicity of zero-eigenvalues for $\text{Dim}(k)$ combinatorial Laplacian matrices during a filtration process. Multiplicities of zero-eigenvalues are equivalent to Betti numbers. Persistent multiplicity is equivalent to Persistent Betti numbers or Betti curves.
	}
	\label{fig:c60_gamma0}
\end{figure}

As an illustration, we consider a PerSpect simplicial complex model of fullerene C$_{60}$. We choose cutoff distance as filtration parameter and use Vietoris-Rips complex to construct simplicial complex. Figure \ref{fig:c60_matrix} demonstrates the generated sequence of nested simplicial complexes and their corresponding combinatorial Laplacian matrixes. Notation $\text{Dim}(k)$ means the $k$-th dimension, and only Laplacian matrixes at $\text{Dim}(0)$ to $\text{Dim}(2)$ are illustrated. It can be seen that, during the filtration process, complexes have been systematically generated and the simplicial complex evolve from a set with isolated vertices to a fully-connected complete topology. The corresponding Laplacian matrixes characterize this evolution process very well. For $\text{Dim}(0)$, at the very start of the filtration, there are only 60 vertices (0-simplex), thus a 60*60 all-zero ${\bf L}_0$ Laplacian matrix is generated. As the increase of filtration value, the size of ${\bf L}_0$ matrix remains unchanged, while more and more -1 value appears at its off-diagonal part. When the filtration value is large enough, a complete graph is obtained, and a full ${\bf L}_0$ matrix, i.e.,  all diagonal entries are 60 and off-diagonal entries are -1, is generated according to Eq.(\ref{eq:Simplex_Laplacian_dim0}). For $\text{Dim}(1)$, at early stage of filtration, there exists no edges (1-simplexes) thus no ${\bf L}_1$ Laplacian matrices. With edges emerging as the filtration value increases, ${\bf L}_1$ matrixes are generated. Different from the $\text{Dim}(0)$ case, the size of ${\bf L}_1$ matrix increases systematically with the number of edges. Off-diagonal entries have both value 1 and -1 due to the orientation of the edge. When the filtration value is large enough, all edges will be either upper adjacent or not lower adjacent, thus ${\bf L}_0$ matrix becomes a diagonal matrix with all its diagonal entry value as 60, according to Eq.(\ref{eq:Simplex_Laplacian_long}). For $\text{Dim}(2)$, no ${\bf L}_2$ Laplacian matrixes exist at the beginning stage of filtration, as no 2-simplexes are generated. The size of ${\bf L}_2$ matrixes also increases with the filtration and eventually evolve into a diagonal matrix with its diagonal entries all as 60 according to Eq.(\ref{eq:Simplex_Laplacian_long}). Mathematically, higher dimensional Laplacian matrixes can also be systematically generated.

\begin{figure}
\begin{center}
\begin{tabular}{c}
\includegraphics[width=0.8\textwidth]{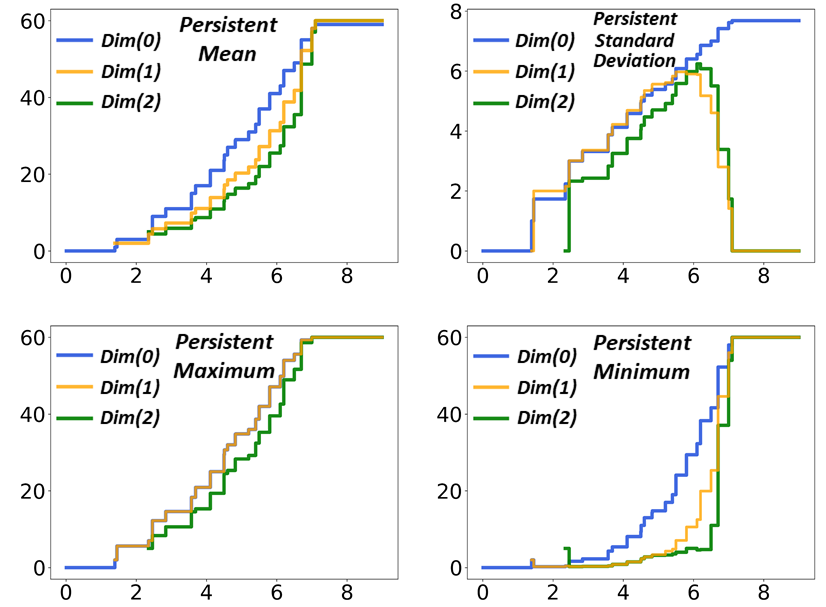}
\end{tabular}
\end{center}
\caption{Illustration of four PerSpect variables, including persistent mean, persistent standard deviation, persistent maximum and persistent minimum, for fullerene C$_{60}$.
}
\label{fig:c60_stat}
\end{figure}

Further, we can study PerSpect variables for fullerene C$_{60}$. Figure \ref{fig:c60_gamma0} shows the comparison between persistent barcode and persistent multiplicity. It can be seen that the persistent multiplicity is equivalent to the persistent Betti function \cite{KLXia:2015a}, defined as the summation of persistent barcodes. In this way, the persistent homology information is naturally embedded into persistent multiplicity. Figure \ref{fig:c60_stat} shows the persistent mean, persistent standard deviation, persistent maximum and persistent minimum for C$_{60}$. It can be seen that these PerSpect variables change with the filtration value. Each variation of PerSpect variables indicates a certain change of the simplicial complex. At filtration size 7.10 \AA, a complete simplicial complex is achieved, i.e., any $k+1$ vertices will form a $k$-simplex. The corresponding ${\bf L}_0$ has eigenvalues 0 (with multiplicity 1) and 60 (with multiplicity 59). The size for the corresponding ${\bf L}_1$ is 1770*1770, and its eigenvalues are all 60 (with multiplicity 1770).  The size for complete corresponding ${\bf L}_2$ is 34220*34220, and its eigenvalues are also 60 (with multiplicity 34220).

\begin{figure*}
	\begin{center}
		\begin{tabular}{c}
			\includegraphics[width=0.95\textwidth]{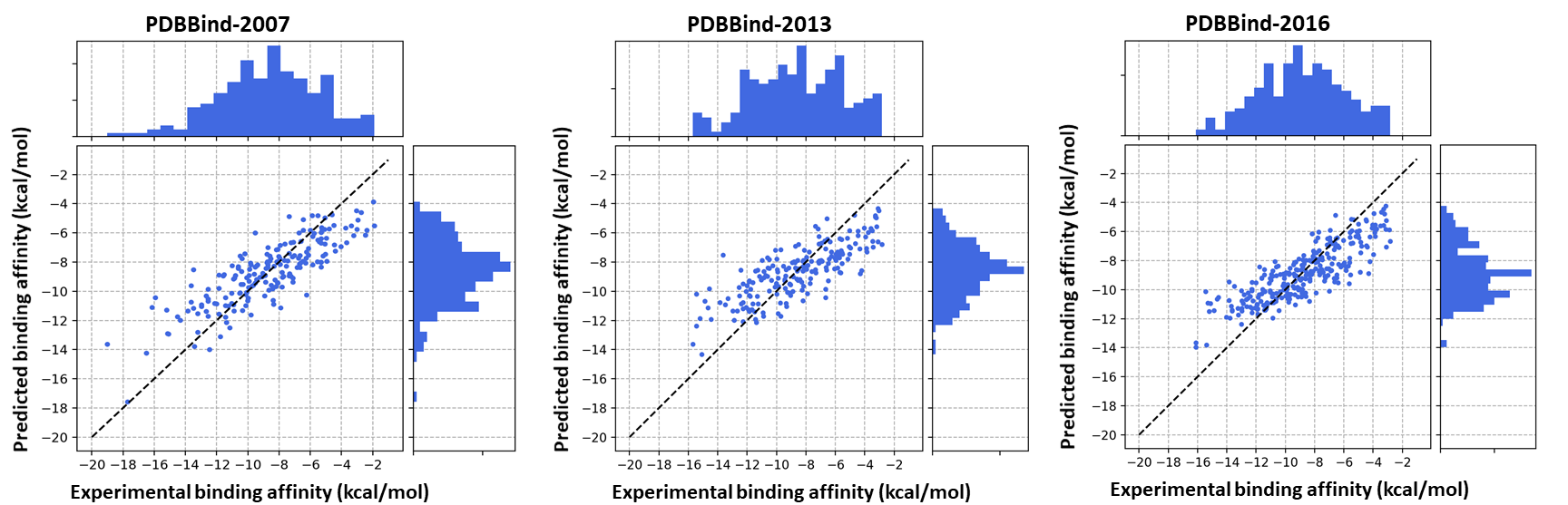}
		\end{tabular}
	\end{center}
	\caption{The comparison of predicted protein-ligand binding affinities and experimental results for PDBbind-2007, PDBbind-2013, and PDBbind-2016. The PCCs are 0.836, 0.793 and 0.840, respectively. The RMSEs are 1.847, 1.956 and 1.724, respectively.
	}
	\label{fig:corr_plot}
\end{figure*}

\subsection{PerSpect ML and its application in drug design}

\paragraph{PerSpect based machine learning models}
Essentially, our PerSpect theory provides a mathematical representation of data, thus can work as a featurization for machine learning models. More specifically, PerSpect variables and functions can be discretized into feature vectors for different learning models. Other than the multiplicity of zero eigenvalues and non-zero eigenvalue statistic properties as mentioned above, spectral indexes from molecular descriptors can also be considered \cite{puzyn2010recent}. For a Laplacian matrix with eigenvalues $ \{\lambda_{1},\lambda_{2},...,\lambda_{n} \}$,  commonly-used spectral indexes include, sum of eigenvalues (Laplacian graph energy), sum of absolute deviation of eigenvalues (generalized average graph energy $\sum_{i=1}^{n}{\left| \lambda_i - \bar{\lambda} \right|}$ with $\bar{\lambda} $ the average eigenvalue), spectral moments ($\sum_{i=1}^{n}{{\lambda_i}^k} $ with $k$ the order of moment), quasi-Wiener index ($ \sum_{j=1}^{A}{\frac{A+1}{\lambda_j}}$ with $\lambda_j > 0$ and $A$ the number of all nonzero eigenvalues), spanning tree number ($\lg{[\frac{1}{A+1}\cdot\prod_{j=1}^{A}{\lambda_j}]}$), etc. Different from previous spectral index based molecular descriptors, which are extracted from a specific graph structure, a series of spectral index, i.e., an index vector, are considered from graphs, simplicial complexes, or hypergraphs obtained from the filtration process. Another potential featurization representation is to construct two dimensional (2D) images from these PerSpect variables \cite{cang:2018representability}. The image representation will be more suitable for deep learning models.

\paragraph{PerSpect models for drug design}
Mathematical representations, that characterize biomolecular structural, physical, chemical and biological properties, are key to the success of machine learning models for drug design. For PerSpect ML based drug design, we consider element-specific (ES) biomolecular topological modeling \cite{cang:2018representability}. Instead of using biomolecular topologies from either all-atom models or coarse-grained models (such as C$_{\alpha}$), we decompose a biomolecule into different point sets, each with only one type of atoms, and construct topological representations for each set. For instance, a protein structure can be decomposed into 5 different points sets, that contain hydrogen(H), carbon(C), nitrogen(N), oxygen(O), and sulfur(S), separatively. Ligands are usually composed of totally 10 types of points sets, with the other 5 types of atoms including phosphorus(P), fluoride(F), chloride(Cl), bromide(Br) and iodine(I), respectively. Our PerSpect models can be employed on each set or each pair of atom sets. In this way, a detailed topological characterization of the biomolecular structure is obtained.

Further, for protein-ligand binding affinity prediction, we consider the interactive distance matrix (IDM) defined as follows \cite{cang:2018representability},
\begin{eqnarray}\label{eq:Distmatrix}
{M}(i,j)=\left\{\begin{array}{ll}\nonumber
               {\scriptstyle \|{\bf r}_i-{\bf r}_j\|},&\text{if~}{\scriptstyle {\bf r}_i\in {\bf R}_P, {\bf r}_j\in {\bf R}_L ~\text{or}~ {\bf r}_i\in {\bf R}_L, {\bf r}_j\in {\bf R}_P} \\
                {\infty},&\text{otherwise}.
                \end{array}
                \right.
\end{eqnarray}
Here ${\bf r}_i$ and ${\bf r}_j$ are coordinates for the $i$- and $j$-th atoms, and $\|{\bf r}_i-{\bf r}_j\|$ is their Euclidean distance. Two sets ${\bf R}_P$ and ${\bf R}_L$ are atom coordinate sets for protein and ligand respectively. Only connections (or interactions) between protein atoms and ligand atoms are considered. Connections between atoms within either protein or ligand are ignored by setting their distance as $\infty$, i.e., an infinity large value. Element-specific interactive distance matrixes (ES-IDM) are considered for protein-ligand binding affinity prediction. That is to say there are totally 4*9=36 types of matrixes between 4 types of atoms from protein, including C, N, O, and S, and 9 types of atoms from ligand, including C, N, O, S, P, F, Cl, Br and I.

To characterize electrostatic properties, the interactive electrostatic matrix (IEM) is defined as follows \cite{cang:2018representability},
\begin{eqnarray}\label{eq:electroDist}
M_E(i,j)=\left\{\begin{array}{ll} \nonumber
\frac{1}{1+\exp({-\frac{cq_{i}q_{j}}{\|{\bf r}_i-{\bf r}_j\|}})},&\text{if~} {\scriptstyle {\bf r}_i\in {\bf R}_P, {\bf r}_j\in {\bf R}_L ~\text{or}~ {\bf r}_i\in {\bf R}_L, {\bf r}_j\in {\bf R}_P} \\
                {\infty},&\text{otherwise}.
\end{array}
\right.
\end{eqnarray}
Here $q_{i}$ and $q_{j}$ are partial charges for the $i$-th and $j$-th atoms, parameter $c$ is constant value. In our calculation $c$ is set to be 100. In this matrixes, electrostatic interactions between atoms within either protein or ligand are dismissed by setting their value as ${\infty}$. Only interactions between protein and ligand are considered. For element-specific interactive electrostatic matrix (ES-IEM), there are totally 5*10=40 types, between 5 types of atoms from protein, including H, C, N, O, and S, and 10 types of atoms from ligand, including H, C, N, O, S, P, F, Cl, Br and I. 

\section{Results and discussions}
In this section, we consider PerSpect simplicial complex based machine learning model for the protein-ligand binding affinity prediction, one of the most important task in drug design.

\subsection{Data preparation}
We choose three most commonly-used protein-ligand databases \cite{PDBBind:2015} for their binding affinity prediction, namely PDBbind-2007, PDBbind-2013 and PDBbind-2016. The data are downloaded from PDBbind (www.pdbbind.org.cn). For each database, the core set is regarded as the test set, all entries in refined set except the ones in the core set form the training set. The detailed data information can be found in Table \ref{tab:database}.

\begin{table}
		\centering
		\caption{Description of the PDBbind databases}
		\begin{tabular}{|c|c|c|c|}  \hline
			Version & Refined set & Training set & Core set (test set)  \\ \hline
			v2007   & 1300        & 1105         & 195        \\ \hline
			v2013   & 1959        & 2764         & 195        \\ \hline
			v2016   & 4057        & 3772         & 285        \\ \hline
		\end{tabular}\label{tab:database}
\end{table}

\begin{figure}
	\begin{center}
		\begin{tabular}{c}
			\includegraphics[width=0.95\textwidth]{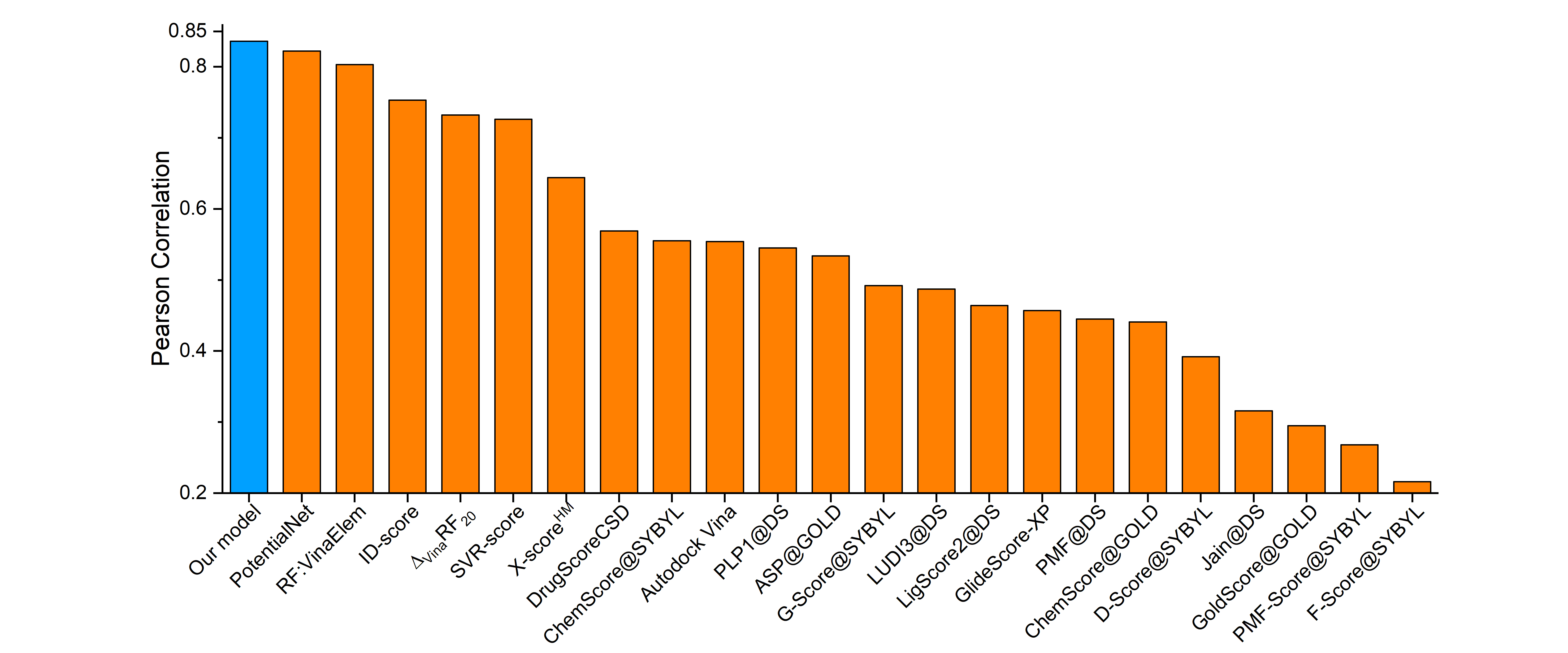}
		\end{tabular}
	\end{center}
	\caption{Performance comparison of our PerSpect simplicial complex based GBT with the-state-of-art models for PDBbind-2007 data \cite{liu2015classification,li2015improving,feinberg2018potentialnet,wojcikowski2019development,jimenez2018k,stepniewska2018development,su2018comparative,zheng2019onionnet,afifi2018improving,feinberg2018potentialnet,boyles2019learning}.
.
	}
	\label{fig:affinity_2007}
\end{figure}

\subsection{Parameter setting}
In our ES-IDM based PerSpect simplicial complex models, the distance value is considered as the filtration parameter. The filtration value goes from 0.00 to 25.00 \AA. For discretization, Laplacian matrixes are generated with a step of 0.10~\AA. That is to say, a totally 250 Laplacian matrixes are generated from each filtration process.  Further, in ES-IEM based PerSpect models, the interaction strength is used as the filtration parameter and its value goes 0.00 to 1.00. The Laplacian matrix is generated with a step of 0.01, meaning totally 100 Laplacian matrixes for each filtration process. Further, we consider 11 PerSpect features, including persistent multiplicity for both $\text{Dim}(0)$ and $\text{Dim}(1)$, persistent mean, persistent standard deviation, persistent maximum, persistent minimum, persistent Laplacian graph energy, persistent generalized mean graph energy, persistent spectral moment (second order), persistent quasi-Wiener index and persistent spanning tree number. Note that other the persistent multiplicity, all other PerSpect variables are calculated only for $\text{Dim}(0)$ Laplacians. To sum up, in our ES-IDMs, there are 36 types of atom combinations as stated above, and the total number of features are 36[types]*250[persistence]*11[eigen feature]. Similarly, there are 50 types of ES-IEMs, and the number of features are 50[types]*100[persistence]*11[eigen feature].

Since we have a large feature vector, decision-tree based models are considered to avoid overfitting. In particular, gradient boost tree (GBT) model have delivered better results in protein-ligand binding affinity prediction. The parameters of GBT are listed in the Table \ref{tab:GBT}. Note that 10 independent regressions are conducted and the medians of 10 PCCs and RMSEs are computed as the performance measurement of our PerSpect ML model.

\begin{table}
  \centering
\caption{The parameters for our GBT model.}
\begin{tabular}{|c|c|c|c|}  \hline
No. of estimators & Max depth & Minimum sample split & Learning rate  \\ \hline
40000             & 6         & 2                    & 0.001         \\ \hline \hline
 Loss function   & Max features      & Subsample size  & Repetition\\ \hline
 least square & square root & 0.7 & 10 times \\ \hline
\end{tabular}\label{tab:GBT}
\end{table}

\begin{table}
  \centering
\caption{The performance of our PerSpect simplicial complex based GBT models in three test cases, i.e., PDBbind-2007, PDBbind-2013 and PDBbind-2016.}
\begin{tabular}{|c|c|c|c|}  \hline
     & ES-IDM  & ES-IEM        & ES-IDM+ES-IEM \\ \hline
PDBbind-2007 & 0.829(1.868) & 0.816(1.941) & 0.836(1.847) \\ \hline
PDBbind-2013 & 0.781(2.005) & 0.786(1.979) & 0.793(1.956) \\ \hline
PDBbind-2016 & 0.830(1.764) & 0.832(1.757) & 0.840(1.724) \\ \hline
\end{tabular}\label{tab:binding_affinity}
\end{table}

\subsection{Basic results}

\begin{figure*}
	\begin{center}
		\begin{tabular}{c}
			\includegraphics[width=0.95\textwidth]{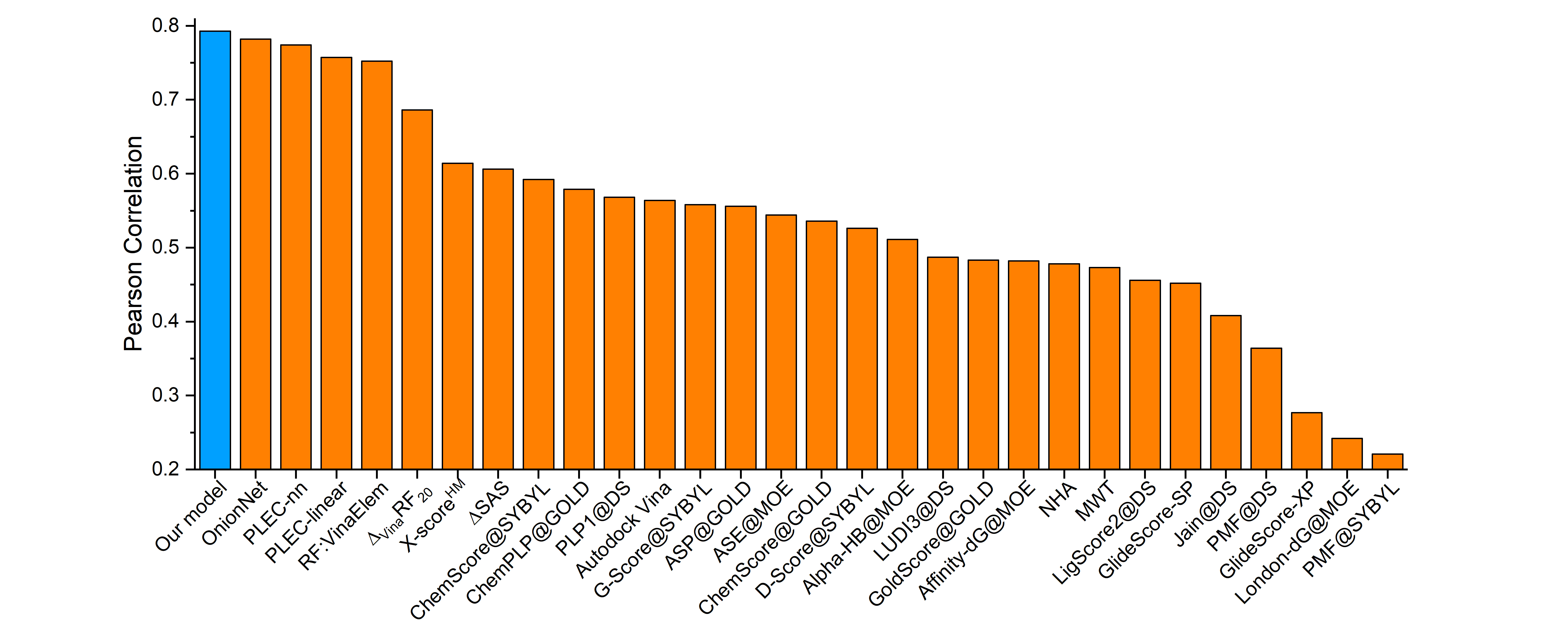}
		\end{tabular}
	\end{center}
	\caption{Performance comparison of our PerSpect simplicial complex based GBT with the-state-of-art models for PDBbind-2013 data \cite{liu2015classification,li2015improving,feinberg2018potentialnet,wojcikowski2019development,jimenez2018k,stepniewska2018development,su2018comparative,zheng2019onionnet,afifi2018improving,feinberg2018potentialnet,boyles2019learning}.
.
	}
	\label{fig:affinity_2013}
\end{figure*}

We consider three GBT models with features from ES-IDM model, ES-IEM model, and combined ES-IDM and ES-IEM mode, respectively. The three models are tested on three databases and an average PCC around 0.800 is obtained. Our PerSpect based GBT results are listed in Table \ref{tab:binding_affinity}. We compare the predicted binding affinity values with the experimental ones, and illustrate the results for the three datasets in Fig. \ref{fig:corr_plot}.

\begin{figure*}
	\begin{center}
		\begin{tabular}{c}
			\includegraphics[width=0.95\textwidth]{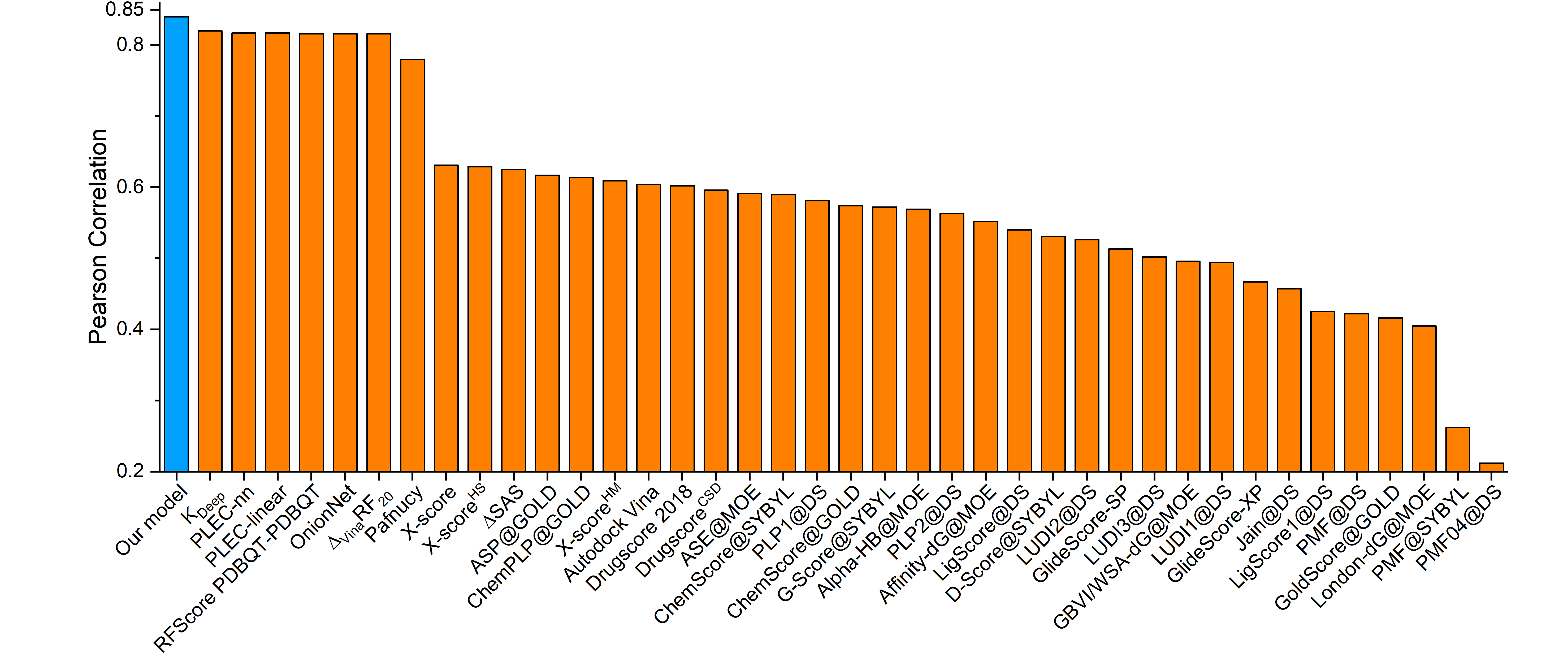}
		\end{tabular}
	\end{center}
	\caption{Performance comparison of our PerSpect simplicial complex based GBT with the-state-of-art models for PDBbind-2016 data \cite{liu2015classification,li2015improving,feinberg2018potentialnet,wojcikowski2019development,jimenez2018k,stepniewska2018development,su2018comparative,zheng2019onionnet,afifi2018improving,feinberg2018potentialnet,boyles2019learning}.
.
	}
	\label{fig:affinity_2016}
\end{figure*}

Further, to have a better understanding of the performance of our models, we have systematically compare our predictions with the state-of-the-art results in literatures \cite{liu2015classification,li2015improving,feinberg2018potentialnet,wojcikowski2019development,jimenez2018k,stepniewska2018development,su2018comparative,zheng2019onionnet,afifi2018improving,feinberg2018potentialnet,boyles2019learning}, as far as we known. The results are illustrated in Figs. \ref{fig:affinity_2007}, \ref{fig:affinity_2013}, and \ref{fig:affinity_2016}. It can be seen that our PerSpect models have achieved the highest PCCs for all three datasets. Further, our RMSE results are also the lowest among all the existing models. This demonstrates the great power of PerSpect theory in biomolecular representation.

\section{Conclusion}
In this paper, we propose persistent spectral models based machine learning (PerSpect ML) for drug design. Three PerSpect models, including persistent spectral graph, persistent spectral simplicial complex and persistent spectral hypergraph, are proposed. A series of persistent spectral variables, including persistent mulitiplicity, persistent mean, persistent maximum, etc, are considered for biomolecular structure characterization and used as features for machine learning models. We systematically test our models on three commonly-used protein-ligand binding databases. Our PerSpect models can achieve the best results for all three datasets.



\section*{Funding}

This work was supported in part by Nanyang Technological University Startup Grant M4081842.110, Singapore Ministry of Education Academic Research fund Tier 1 RG31/18 and Tier 2 MOE2018-T2-1-033. 

\end{document}